\documentclass[twocolumn,showpacs,floats,floatfix,superscriptaddress,aps,pra]{revtex4-1}
\usepackage{amsfonts}
\usepackage{amssymb}
\usepackage{amsmath}
\usepackage{calc}
\usepackage{graphicx}
\usepackage{bm}

\usepackage[normalem]{ulem}
\usepackage{amsmath,amssymb}
\usepackage{multirow}
\usepackage{xcolor,soul}
\usepackage{srcltx}
\usepackage{hyperref,graphicx}

\def\be{ \begin{equation}}
\def\ee{ \end{equation}}
\def\bea{ \begin{eqnarray}}
\def\eea{ \end{eqnarray}}
\def\bse{ \begin{subequations}}
\def\ese{ \end{subequations}}
\def\bc{ \begin{center}}
\def\ec{ \end{center}}

\begin{document}

\author{Stefano Longhi}
\email{stefano.longhi@polimi.it}
\affiliation{Dipartimento di Fisica, Politecnico di Milano and Istituto di Fotonica e Nanotecnologie del Consiglio Nazionale delle Ricerche, Piazza L. da Vinci 32, I-20133 Milano, Italy}

\title{Non-Hermitian quantum rings}
\date{\today }

\begin{abstract}
We investigate  the spectral and dynamical properties of a quantum particle constrained on a ring threaded by a magnetic flux
 in presence of a complex (non-Hermitian) potential. For a static magnetic flux, the quantum states of
the particle on the ring can be mapped into the Bloch states of a complex crystal, and magnetic flux tuning enables to probe the spectral features of the 
complex crystal, including the appearance of exceptional points. For a time-varying (linearly-ramped) magnetic flux,  Zener tunneling among energy states is realized owing to the induced electromotive force. As compared to the Hermitian case, striking effects are observed in the non-Hermitian case, such as  a highly asymmetric behavior of particle motion when reversing the direction of the magnetic flux and field-induced delayed transparency. 

\end{abstract}

\maketitle

\section{Introduction}

The coherent motion of charge
carriers in doubly connected (ring) topologies plays a fundamental role in quantum and mesoscopic physics. Quantum mechanical experiments in ring geometries have long
fascinated physicists.
For example, the quantum orbital motion of electrons in mesoscopic
normal-metal rings threaded by a magnetic fiux produces
striking interference phenomena such as the Aharonov-Bohm effect \cite{R1} and persistent currents \cite{R2,R3,R4}. 
Experimental evidence for Aharonov-Bohm
oscillations has been detected in the mesoscopic regime in
metallic \cite{R5,R6} and semiconducting \cite{R7,R8} rings.  Because
of the periodic boundary conditions enforced by the single
valuedness of the wave function, the eigenstates of
the electron in a ring look like Bloch waves in a
crystal, where the circumference of the ring corresponds to the lattice constant
and the Bloch wave number (quasi-momentum)  in the crystal is taken
up by the flux parameter 
 \cite{R4,R9}.
  The effect of a superimposed nonuniform potential $V(\varphi)$ in the ring simulates the structure of a crystal with allowed and forbidden energy bands and gaps, among which Zener transitions can be induced by the electromotive force created by a linear variation in time of the the magnetic flux \cite{R9}. Such previous studies on quantum rings have been mostly limited to consider an underlying Hermitian Hamiltonian. A noticeable exception is provided by the works by Hatano and Nelson \cite{Hatano}, who investigated non-Hermitian localization in a random Schr\"{o}dinger equation subjected to a constant {\it imaginary vector potential}. \par 
 In this work we study the spectral and dynamical properties of a quantum particle on a ring in the {\it non-Hermitian} case by allowing the {\it external potential} $V(\varphi)$ (rather than the vector potential) to be complex-valued.  Non-Hermitian quantum mechanics has received an increasing interest in recent years \cite{Moiseyev}, especially in the context of Hamiltonians showing space-time reflection ($\mathcal{PT}$) symmetry  \cite{Bender_PRL_98,Bender_RPP_2007,Mos1}. $\mathcal{PT}$ Hamiltonians admit of an entirely real-valued energy spectrum below a phase transition (symmetry-breaking) point, above which pairs of complex-conjugate energies appear \cite{Bender_RPP_2007}. An important class of non-Hermitian systems is provided by complex periodic potentials \cite{P1,P2,P3,P4,P4bis,P5}, which realize a kind of synthetic complex crystals. As compared to ordinary crystals,  complex crystals show unusual scattering and transport properties, which have been investigated in several recent works \cite{P2,P3,P4,P4bis,P5,P6,P7}. Complex crystals have been experimentally
realized in different  physical systems, including open two-level atomic systems interacting with near resonant
light \cite{P2} and optical structures with gain and loss regions \cite{P6,P7}. Here we show that a quantum particle on a ring threaded by a magnetic flux
 in presence of a complex (non-Hermitian) potential $V$ behaves like a Bloch particle in a complex crystal, where the magnetic flux determines the Bloch wave number of the particle. In particular, by tuning the magnetic flux the full spectral band structure of the equivalent complex crystal can be probed, including the onset of spectral singularities \cite{P4bis,spec} and the transition from a real to a complex energy spectrum.  Striking effects are found for non-stationary (linearly-increasing) magnetic fields, where multilevel Landau-Zener (LZ) transitions arise owing to the induced electromotive force. Analytical and numerical results are presented for a particle in the complex potential $V(\varphi)=V_0 \cos (\varphi)+V_0 \alpha \sin (\varphi)$ threaded by a linearly-increasing magnetic flux $f= \beta t$, where LZ transitions occur among quantum states with different winding  (angular momentum) numbers. 
As compared to LZ transition in the Hermitian case ($\alpha=0$),  striking effects are found in the non-Hermitian case ($\alpha \neq 0$), including strong asymmetric behavior for reversal the direction of the magnetic flux and field-induced delayed transparency at the $\mathcal{PT}$ symmetry-breaking transition.

 \section{Quantum particle on a ring threaded by a magnetic flux}
 \subsection{The model}
   The time-dependent Schr\"{o}dinger equation for an electron of mass $m$ and charge $e$ moving on a ring of radius $R$ in presence of the external potential $V(\varphi)$ and threaded by a magnetic flux $\phi=\phi(t)$ [see Fig.1(a)] reads
 \begin{equation}
 i \hbar \frac{\partial \psi}{\partial t}= \frac{\hbar^2}{2mR^2} \left( -i \frac{\partial}{\partial \varphi}-f \right)^2 \psi+V(\varphi) \psi \equiv \hat{H}(t) \psi
 \end{equation} 
 where $ \varphi$ is the azimuthal angle that measures the position of the electron on the ring , $f= \phi/ \phi_0$, and $\phi_0=hc/e$ is the flux quantum. The ring boundary condition 
 \begin{equation}
 \psi(\varphi+ 2 \pi,t)=\psi( \varphi,t)
 \end{equation}
 applies to the single-valued wave function $\psi$. For a time-independent magnetic flux, Eq.(1) can be reduced to a standard one-dimensional Schr\"{o}dinger equation after the gauge transformation $\psi(\varphi,t)=F(\varphi,t) \exp(i f \varphi)$, which simplifies Eq.(1) into the following one
 \begin{equation}
 i \hbar \frac{\partial F}{\partial t}= -\frac{\hbar^2}{2mR^2} \frac{\partial^2 F}{\partial \varphi^2} +V(\varphi) F.
 \end{equation} 
 The magnetic flux controls the boundary condition for the function $F$,  namely one has
 \begin{equation}
 F( \varphi+ 2 \pi,t)=F( \varphi,t) \exp(-2 \pi if)
 \end{equation}
 where the additional phase term on the right hand side of Eq.(4) is the Aharonov-Bohm phase.\par
The Schr\"{o}dinger equation (1) is usually introduced to describe the coherent electronic motion in  mesoscopic metal rings, however it can be found  in other physical contexts as well, where the introduction of a complex-valued external potentials $V(\varphi)$ might be feasible. For example, Eq.(1) can describe the temporal dynamics of  a dilute and rotating Bose-Einstein condensate in an annular trap \cite{BEC},  or spatial propagation of monochromatic light waves in an annular fiber with a twisted axis \cite{fibra}. In the latter optical system, a complex potential $V(\varphi)$ describes the effects of an azimuthal index (real part of $V$) and loss/gain (imaginary part of $V$) guiding. 
\begin{figure}[t]
\includegraphics[width=8cm]{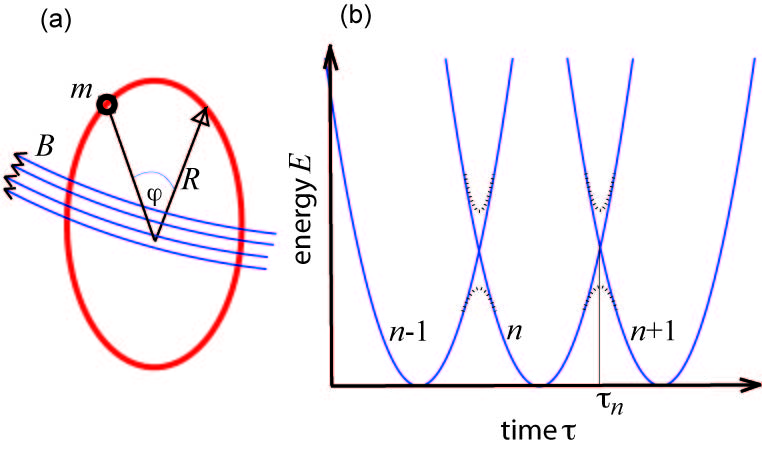}
\caption{(Color online). (a) Schematic of a quantum particle of mass $m$ moving on a ring of radius $R$ threaded by a magnetic field $B=B(t)$. (b) Instantaneous energy levels versus normalized time $\tau$ for the level chain described by Eqs.(19). The solid lines are the diabatic energy levels, corresponding to the sequence of parabola $E=E_n(\tau)=(n - \sigma \tau)^2$, where $n=0, \pm 1 , \pm 2, ...$ is the winding number. The adjacent diabatic levels $n$ and $n+1$ cross at time $\tau_n$ defined by Eq.(21). The adiabatic energy levels are shown by the dotted curves for $\alpha < \alpha_c=1$. The adiabatic levels are almost overlapped with the diabatic ones, except near the crossing times $\tau_n$ where diabatic crossings are transformed into avoided crossings. At the $\mathcal{PT}$ symmetry breaking point ($\alpha=\alpha_c=1$), the adiabatic energy levels coincide with the diabatic ones.}
\end{figure}
 \subsection{Energy spectrum in a static magnetic flux}
 
 In the absence of the external potential, $V(\varphi)=0$, and for a stationary magnetic flux the eigenstates and corresponding energies of the Hamiltonian $\hat{H}$ are given by
 \begin{eqnarray}
 \psi_n(\varphi) & = &  \frac{1}{\sqrt {2 \pi}} \exp(i n \varphi)\\
 E_n & = & \frac{\hbar^2(n-f)^2}{2mR^2}
 \end{eqnarray}
 where $n=0, \pm 1 , \pm 2 , \pm 3, ...$ is the winding number that determines the angular momentum $n \hbar$ of the rotating electron. In the presence of the external potential $V( \varphi)$, the energy spectrum and corresponding eigenfunctions of $\hat{H}$ can be mapped into the spectrum and Bloch-Floquet states of the associated periodic potential $V(\varphi+2 \pi)=V(\varphi)$, where the normalized magnetic flux $f$ plays the role of the wave number $k$ of the electron in the crystal \cite{R9}. To show such an equivalence, let us look for a solution to the eigenvalue problem $\hat{H} \psi (\varphi)=E \psi(\varphi)$ of the form $\psi(\varphi)=F(\varphi) \exp(if \varphi)$. The function $F(\varphi)$ then satisfies the stationary Schr\"{o}dinger equation
 \begin{equation}
 -\frac{\hbar^2}{2mR^2} \frac{d^2 F}{d \varphi^2}+V(\varphi)F=E F
 \end{equation}
 with the boundary condition 
 \begin{equation}
 F(\varphi+2 \pi)=F(\varphi) \exp(-2 \pi i f).
 \end{equation} 
  Since $V(\varphi+2 \pi)=V(\varphi)$, Eq.(7) can be viewed as the stationary Schr\"{o}dinger equation of an electron in the periodic potential $V(\varphi)$ with lattice constant $a=2 \pi$. According to the Bloch-Floquet theorem, the most general solution to Eq.(7) is a Bloch state, $F(\varphi)=u_n(\varphi,k)\exp(ik \varphi)$ and $E=E_n(k)$, where $k$ is the crystal wave number (quasi-momentum), that varies in the interval $-1/2 \leq k < 1/2$, $E_n(k)$ is the energy dispersion curve of the $n$-th band of the crystal, and $u_n(\varphi+2 \pi,k)=u_n(\varphi,k)$ is the periodic part of the Bloch eigenfunction. The boundary condition (8) requires $k=-f$, i.e. the normalized magnetic flux $f$ fixes the quasi momentum $k$ of the electron in the lattice. The eigenfunctions and corresponding energies of the quantum-ring Hamiltonian $\hat{H}$ are thus given by $\psi_n(\varphi)=u_n(\varphi,-f)$ and $E_n=E_n(-f)$. 
 Hence by tuning the normalized magnetic flux from $f=-1/2$ to $f=1/2$ one can probe the entire band structure of the periodic potential $V(\varphi)$.\par For a complex potential, we can generally write
 \begin{equation}
 V(\varphi)=V_R(\varphi)+i \alpha V_I(\varphi),
 \end{equation}
 where $V_R(\varphi)$, $ \alpha V_I (\varphi)$ are the real and imaginary parts of $V(\varphi)$, respectively, $\alpha \geq 0$ is a real parameter that determines the strength of non-Hermiticity of the potential, and $V_R(\varphi)$, $V_I(\varphi)$ are the profiles of the real and imaginary potential terms.
Note that $\alpha=0$ corresponds to the ordinary Hermitian problem. Of particular interest is the case of a $\mathcal{PT}$ symmetric complex crystal, which requires $V(-\varphi)=V^*(\varphi)$.  In this case, a critical value $\alpha_c$ of $\alpha$ does exist such that the energy spectrum of the crystal is entire real-valued for $\alpha \leq \alpha_c$ (unbroken $\mathcal{PT}$ phase), whereas complex-conjugate energies appear for $\alpha> \alpha_c$ (broken $\mathcal{PT}$ phase). The complex-conjugate energies above the symmetry breaking point emanate from the wave numbers $k=0$ or $k= \pm \pi$, i.e. at the center or at the edge of the Brillouin zone \cite{P4bis}. For example, for the potential
 \begin{equation}
 V(\varphi)=V_0 \cos(\varphi)+i \alpha V_0 \sin (\varphi)
 \end{equation}
 one has $\alpha_c=1$ \cite{P4}. Interestingly, at the $\mathcal{PT}$ symmetry breaking point spectral singularities, corresponding to poles in the resolvent
of the Hamiltonian in the continuous part of the spectrum, are found at either the center ($k=0$) or at the edge ($k= \pm 1/2$) of the Brillouin zone \cite{P4bis}.  The physical implications of spectral singularities in non-Hermitian systems have been highlighted in recent works, especially in connection to resonance-like behavior in scattering problems and instability thresholds in optical systems (see, for instance, \cite{spec}). 
 In particular, in complex crystals spectral singularities  at the $\mathcal{PT}$ symmetry breaking point are responsible for a secular growth in time of the wave function $\psi(\varphi,t)$ in spite of the real energy spectrum of the Hamiltonian \cite{P4bis}. Mathematical implications of spectral singularities have been investigated as well \cite{spec,specM1,specM2,specM3}, and the question whether resolution of the identity operator is possible for a Hamiltonian
possessing spectral singularities  has been debated. To this regard, contrary to earlier indications it was rigorously proven  in Ref.\cite{specM2} that the contribution of the spectral singularity to the resolution
of identity operator depends on the class of functions employed for physical states, and that there is no obstruction to completeness originating from a spectral singularity (see also Ref.\cite{specM3}). 
In case of a non-Hermitian quantum ring considered in this paper, it should be nevertheless noticed that the spectrum of the Hamiltonian is point like, 
and spectral singularities  of the associated complex crystal are mapped into {\it exceptional points} \cite{Moiseyev,EP1,EP2,EP3}.  Thus, for the quantum particle on the ring threaded by a magnetic flux with the external potential $V(\varphi)$ at $\alpha=\alpha_c$, the eigenfunctions of $\hat{H}$ form a complete basis for a  magnetic flux $f$ different than either (or both) $f=n$, $f= \pm 1/2+n$ ($n=0, \pm 1, \pm 2, ...$), whereas in the opposite case coalescence of pairs of eigenfunctions and eigenvectors (corresponding to exceptional points) are found. \par 
 As an example, let us consider the periodic potential (10) at $\alpha=\alpha_c=1$, i.e. 
 \begin{equation}
 V(\varphi)=V_0 \exp(i \varphi).
 \end{equation}
  The energy spectrum and corresponding eigenfunctions of $\hat{H}$ can be calculated in a closed form  (see, for instance, \cite{P1,P4bis}). In particular, the energy spectrum  turns out to be the same as that of a free particle, i.e.
$ E_n$ is given by Eq.(6). For $f \neq (2l+1)/2$ (with $l=0, \pm 1, \pm 2, \pm 3, ...$) the eigenvalues of $\hat{H}$ are simple (non-degenerate) and the corresponding eigenfunctions form a complete set. As $f$ approaches a value close to half an integer, i.e. as $f \rightarrow (2l+1)/2$, two enenergies coalesce in pairs, namely $E_{l+1} - E_{l} \rightarrow 0$, and the corresponding set of eigenfunctions ceases to be complete because of the exceptional point at $E=E_{l}$. The appearance of the exceptional point leads to a secular growth in time of an initial wave function with a defined winding number. In fact, let us expand the wave function $\psi(\varphi,t)$ on the basis of functions with defined winding number, defined by Eq.(5), i.e. let us set
\begin{equation}
\psi(\varphi,t)=\frac{1}{\sqrt{2 \pi}}\sum_{n=-\infty}^{\infty} c_n(t) \exp(in \varphi-iE_nt)
\end{equation}
where $E_n$ are given by Eq.(6).
After substitution of Eq.(12) into Eq.(1) and assuming the potential (11), the following evolution equations for the amplitude probabilities $c_n(t)$ are readily found
\begin{equation}
i \hbar \frac{dc_n}{dt}=V_0c_{n-1} \exp[i(E_n-E_{n-1})t].
\end{equation}
Let us assume that the particle is initially prepared in a state with  a definite angular momentum, corresponding to the winding number $n=n_0$, i.e. that $c_{n}(0)=\delta_{n,n_0}$. The the solution to the coupled equations (13) can be derived from the following recursive relations
\begin{eqnarray}
c_n(t) & = & 0 \; \; (n <n_0) \nonumber \\
c_{n_0}(t) & = & 1 \\
c_{n}(t) & = & - \frac{iV_0}{\hbar} \int_0^t d \xi c_{n-1}(\xi) \exp[i(2n-2f-1) \xi] \; \; (n>n_0) \;\;\; \nonumber 
\end{eqnarray}
From Eqs.(14) it follows that the solution $c_{l+1}(t)$ for $l \geq n_0$ secularly grows in time provided that $2l-2f+1=0$, which is satisfied for a normalized magnetic flux $f$ 
 given by $f=(2l+1)/2$. Hence tuning the magnetic flux at an exceptional point leads to a secular growth in time of the wave function.\par
  As a final comment, it should be noted that, while there is a close connection between the problem of a quantum particle on a ring and the related complex crystal problem, the evolution of a quantum wave packet in the two cases can show rather distinctive features as a result of the restricted spectrum of $\hat{H}$ in the quantum ring problem. For example, let us consider the complex potential $V(\varphi)=V_0 \exp(i \varphi)$ at the $\mathcal{PT}$ symmetry breaking point. In a complex crystal, the spectral singularities of $\hat{H}$ are responsible for an initial growth a normalizable state (a wave packet), however the wave function growth is limited because of the zero measure of the spectral singularities, as discussed in Ref.\cite{P4bis}. Contrary,  
  in the quantum ring problem the growth is not clamped. The reason thereof is that the spectral singularities are transformed into exceptional points in the quantum ring problem, which belong to the point spectrum of $\hat{H}$. 
 \begin{figure}[t]
\includegraphics[width=8.7cm]{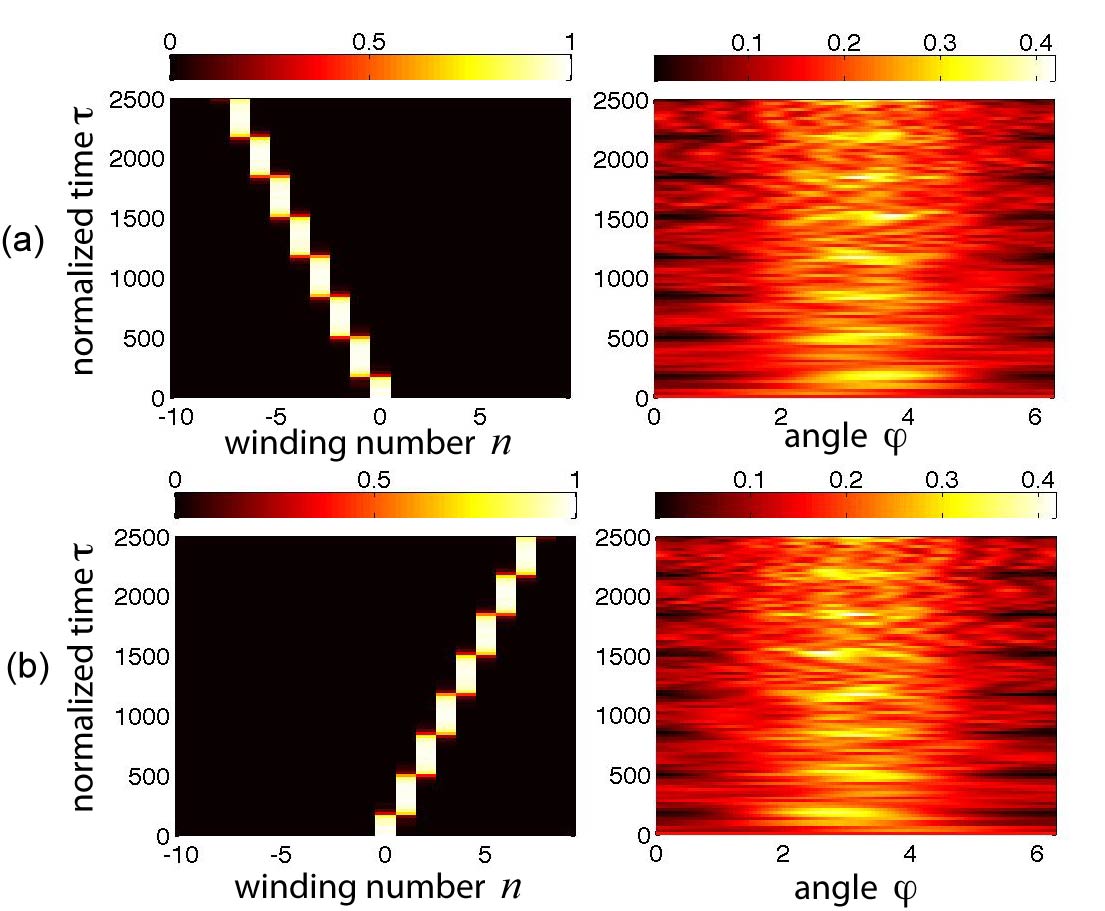}
\caption{(Color online). Evolution of a quantum particle on a ring threaded by a linearly-varying magnetic flux in momentum space [left panels, snapshot of $|c_n(\tau)|^2$ ] and in real space [right panels, snapshot of $|\psi(\varphi,\tau)|^2$]  in the Hermitian case ($\alpha=0$) for (a)  $\sigma=-0.003$, and (b) $\sigma=0.003$. The particle is initially at rest and fully delocalized in the ring, corresponding to $\psi(\varphi,0)=1/ \sqrt{2 \pi}$. The amplitude of the external potential is $V_0mR^2/ \hbar^2=0.08$.}
\end{figure}
 \begin{figure}[t]
\includegraphics[width=8.7cm]{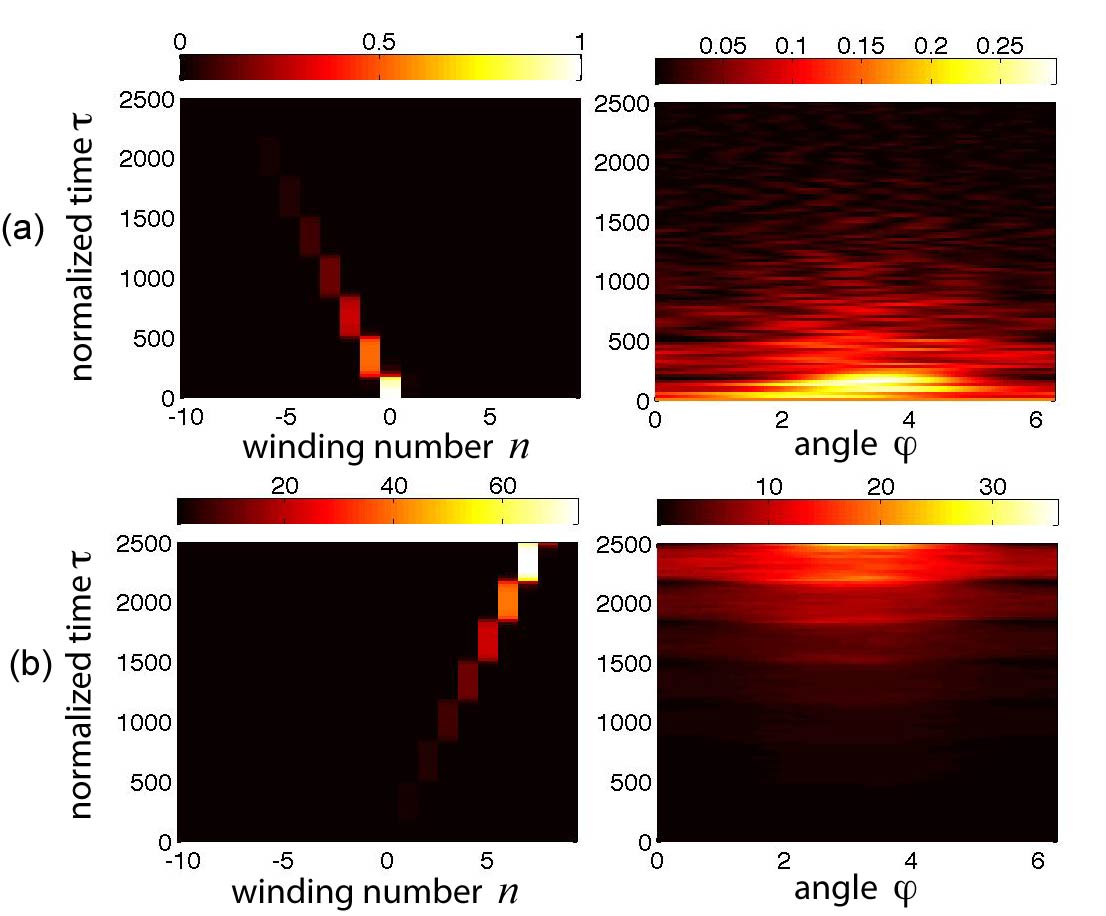}
\caption{(Color online). Same as Fig.2, but in the non-Hermitiain case below the $\mathcal{PT}$-symmetry breaking point $(\alpha=0.3)$. }
\end{figure}
 \subsection{Time-varying magnetic flux: Zener transitions}
 A particularly interesting case is the one of a magnetic flux which is varied in time \cite{R9,uffa}.  When the flux $\phi(t)$ threading the ring is linearly increased in time, i.e.
 \begin{equation}
 f(t)=\beta t
 \end{equation}
 a constant electric field (electromotive force) which accelerates the particle is induced in the ring according to Faraday's law  \cite{note}. 
  Since $k=-f$, a linearly-varying magnetic flux $f=\beta t$ corresponds to a Bloch electron moving at a constant speed $\beta$ in the momentum ($k$) space.  An electron
in a pure Bloch state will follow the flux change adiabatically
if the induced electric field is infinitesimal, i.e., it
will be backscattered to the same energy band each time
it reaches a zone boundary. If the field strength [i.e. the $\beta$ parameter in Eq.(15)] is increased,
 Zener tunneling between different bands can occur. For a real-valued potential $V(\varphi)$, i.e. in the Hermitian case, the general problem of 
 Zener tunneling of a quantum particle on a ring threaded by a ramped magnetic flux can be investigated rather generally  by decomposing the wave function $\psi(\varphi,t)$ as a superposition of the adiabatic Bloch states, namely $\psi(\varphi,t)=\sum_n a_n(t) u_{n}(\varphi,k=-\beta t)$, and looking for the evolution of the amplitudes $a_n(t)$ (see, for instance, \cite{R9}). Zener tunneling between adjacent bands is ruled by a cascade of two-level LZ tunneling  events, which occur as the magnetic flux $f$ crosses the edge of the Brillouin zone. While such a method can be extended to the non-Hermitian case \cite{P5}, it has a limited validity since it fails in the presence of exceptional points. The reason thereof is that, as magnetic flux $f(t)$ crosses an exceptional point, the adiabatic eigenstates of $\hat{H}$ lack of completeness. Therefore, in the non-Hermitian case it is more convenient to  study the quantum evolution by expanding the wave function $\psi(\varphi,t)$ on the basis of the angular momentum eigenfunctions (5). After setting
 \begin{equation}
 \psi(\varphi,t)=\frac{1}{\sqrt{2 \pi}} \sum_{n=-\infty}^{\infty} c_n(t) \exp(i n \varphi)
 \end{equation}
  and 
  \begin{equation}
  V(\varphi)=\sum_{n=-\infty}^{\infty} V_n \exp(i n \varphi),
  \end{equation}
   substitution of Eqs.(16) and (17) into Eq.(1) yields the following evolution equations for the amplitudes $c_n(t)$
   \begin{equation}
   i \hbar \frac{dc_n}{dt}=\frac{\hbar^2}{2mR^2} (n-\beta t )^2c_n+\sum_{m}V_{n-m}c_m.
   \end{equation}
 In their present form, Eqs.(18) can be regraded as a multi-level LZ problem \cite{ML}, where successive crossings of the bare (diabatic) energy levels between states $n$ and $m$ occurs at the times $t_{n,m}=(n+m)/(2 \beta)$.  
 
 \section{Multilevel non-Hermitian Landau-Zener transitions and field-induced delayed transparency}
In this section we focus our attention to the multi-level LZ problem [Eqs.(18)] for the specific potential given by Eq.(10) with $\alpha \leq \alpha_c=1$, and highlight distinct features of non-Hermitian versus Hermitian case. In particular, striking effects are predicted in the non-Hermitian case at the $\mathcal{PT}$ symmetry breaking point, as discussed below.  For the potential (10), after introduction of the normalized time $\tau=\hbar t/(2mR^2)$, Eqs.(18) read
   \begin{equation}
   i  \frac{dc_n}{d \tau}=(n-\sigma \tau )^2c_n+S_1c_{n-1}+S_2 c_{n+1}
   \end{equation}
where we have set
\begin{equation}
S_1=\frac{ V_0 m R^2}{\hbar^2}(1+ \alpha) \; , \; \; S_2=\frac{V_0 m R^2}{\hbar^2}(1-\alpha)
\end{equation}
and $\sigma=2 m R^2 \beta / \hbar$. Note that $S_1=S_2$ in the Hermitian case  ($\alpha=0$), whereas $S_1 \neq 0$, $S_2=0$ at the $\mathcal{PT}$ symmetry breaking point ($\alpha=\alpha_c=1$). The coupled equations (19) can be regarded as a generalization, to the non-Hermitian case, of a multi-level LZ problem \cite{NH2L,Morales}. 
A particularly interesting case is that of a shallow potential, corresponding to $V_0 \ll \hbar^2/(2mR^2)$ (i.e. $S_{1,2} \ll 1$), and a slow increase of the magnetic flux, $\beta \ll  \hbar/(2mR^2)$
(i.e. $| \sigma| \ll 1$), with $S_{1,2} / \sqrt{\sigma}$ of the order (or larger than) $ \sim 1$. 
 In this case, the multilevel LZ problem reduces to the cascade of Zener transitions between two levels $n$ and $(n+1)$,  as one can see from the energy level diagram  corresponding to Eqs.(19) and shown in Fig.1(b). Since $S_{1,2} \ll 1$, the diabatic energies of the levels are far apart each other that transitions are not allowed, except in the neighborhood of  the times 
 \begin{equation}
 \tau_n=\frac{2n+1}{2 \sigma}
\end{equation} 
$(n=0,1,2,3,...)$, where crossing of the energies between level $n$ and level $(n+1)$ occurs [see Fig.1(b)].  This means that, apart from the dynamical phase term, the amplitude $c_n(t)$ does not change in time, except for sudden changes at the two crossings times $\tau_{n-1}$ and $\tau_{n}$. For example, at $\tau \sim \tau_n$ the change of amplitudes $c_n$ and $c_{n+1}$ can be obtained by solving the two-level LZ problem
 \begin{eqnarray}
 i \frac{dc_{n}}{d \tau} & = & (n-\sigma \tau)^2 c_{n}+S_{2}c_{n+1} \\
 i \frac{dc_{n+1}}{d \tau} & = & (n+1-\sigma \tau)^2 c_{n+1}+S_{1}c_{n}
 \end{eqnarray}
 which shows a linear level crossing at $\tau=\tau_n$. The scattering matrix, that relates the values of $c_n$, $c_{n+1}$ at times $\tau \ll \tau_n$ and $\tau \gg \tau_n$, does not depend on the winding number $n$ and its form can be rather generally expressed in terms of parabolic cylinder functions \cite{Vita}.
 It should be noted that, below the $\mathcal{PT}$ symmetry breaking point, i.e. for $\alpha<1$, the non-Hermitian multilevel LZ problem (19) can be readily mapped into the corresponding Hermitian one. In fact, after introduction of the amplitudes $a_n(t)$ by the relation
 \begin{equation}
 c_n(t)=a_n(t)\left( \frac{1+\alpha}{1-\alpha} \right)^{n/2}
 \end{equation}
 Eqs.(19) take the form
 \begin{equation}
   i  \frac{da_n}{d \tau}=(n-\sigma \tau )^2a_n+S(a_{n-1}+ a_{n+1})
   \end{equation}
 where we have set $S \equiv S_1[(1-\alpha)/(1+\alpha)]^{1/2}=S_2 [(1+\alpha)/(1-\alpha)]^{1/2}$. In their present form, Eqs.(25) describe the multilevel LZ problem (19) in the Hermitian case, with $S_1=S_2=S$. Hence the evolution for the amplitudes $c_n(t)$ in the non-Hermitian case, below the $\mathcal{PT}$ phase transition, can be readily obtained from the  behavior $a_n(t)$ of the associated Hermitian problem after the substitution defined by Eq.(24). The major effect of non-Hermiticity in the LZ problem is the breakdown of the time reversal symmetry. This implies an asymmetric behavior of the particle motion when the sign of the magnetic flux, i.e. the direction of the magnetic field threading the ring, is reversed. In fact, let us first observe that, if $a_n(t,\sigma)$ is a solution to Eq.(25) with a magnetic flux $f=\sigma \tau$, then one has $a_n(t,-\sigma)=a_{-n}(t, \sigma)$, i.e. reversal of the direction of the magnetic field (and hence of the electromotive force) in the Hermitian case merely corresponds to reverse the direction of motion on the ring ($n \rightarrow -n$). On the other hand, for the non-Hermitian case, from Eq.(24) it follows that 
 \begin{eqnarray}
 c_n(t,\sigma) & = & a_n(t,\sigma) \left( \frac{1+\alpha}{1-\alpha} \right)^{n/2} \\
 c_n(t,-\sigma) & = & a_{-n}(t,\sigma) \left( \frac{1+\alpha}{1-\alpha} \right)^{n/2}  
 \end{eqnarray}
 and hence:
 \begin{equation}
 c_n(t,-\sigma)  = c_{-n}(t,\sigma)  \left( \frac{1+\alpha}{1-\alpha} \right)^{n}
 \end{equation}
 i.e. the invariance of the dynamics to the transformations $\sigma \rightarrow -\sigma$, $n \rightarrow -n$ is broken. As an example, let us consider the dynamics of a quantum particle that is initially at rest and fully delocalized on the ring, i.e. let us assume $c_n(0)=\delta_{n,0}$.  In Figs.2 and 3 we show the numerically-computed quantum evolution of the particle state, both in the physical and momentum space, in the Hermitian (Fig.2) and non-Hermitian (Fig.3) case and for two opposite values of the magnetic flux rate $\sigma$. Parameter values have been chosen such that, at each LZ crossing, the probability of Zener tunneling from one level to the coupled one, given by $P_Z=1-\exp(-\pi S^2 / \sigma)$,  is close to one. Hence, in the Hermitian case the effect of the linearly-increasing magnetic flux is to induce a drift of the particle motion in momentum space; the direction of the drift is reversed as the sign of the magnetic flux is flipped, as shown in Fig.2. In the non-Hermitian case, a similar behavior is observed, however the norm of the wave function is not conserved; according to Eq.(28), amplification or damping of the wave function is observed, depending on the sign of $\sigma$ (see Fig.3).\par More striking features can be observed in the non-Hermitian case at the $\mathcal{PT}$ symmetry breaking point $\alpha=\alpha_c=1$. In this case, the non-Hermitian problem can not be mapped into the Hermitian one by the transformation (24) owing to a divergence at $\alpha=1$, and a direct analysis of Eqs.(19) with $S_2=0$ should be considered. In the limit $| \sigma | \ll 1$, $|S_1| \ll 1$ and $S_1/ \sqrt{ | \sigma |}$ larger than (or of the order of) $\sim 1$ , a simple analytical expression to the solution to Eqs.(19) can be derived by an asymptotic analysis. 
 After setting
 \begin{equation}
c_n(\tau)  = a_n(\tau)  \exp \left[ -i \int_0^\tau dt (n-\sigma t)^2\right] 
 \end{equation}
 for $\tau \geq0$ one obtains
 \begin{widetext}
  \begin{equation}
a_n(\tau) \simeq  
\left\{
\begin{array}{ll}
a_n(0) & n \leq 0 \\
 a_n(0)-i S_1 \sqrt{\frac{\pi}{i \sigma}} \exp(i \sigma \tau_{n-1}^2) a_{n-1}(\tau_{n-1}) H(\tau-\tau_{n-1}) & n \geq 1
 \end{array}
  \right.
  \end{equation}
 for $\sigma>0$, whereas 
  \begin{equation}
a_n(\tau) \simeq  
\left\{
\begin{array}{ll}
a_n(0) & n \ge 1 \\
 a_n(0)-i S_1 \sqrt{\frac{\pi}{i \sigma}} \exp(i \sigma \tau_{n-1}^2) a_{n-1}(0) H(\tau-\tau_{n-1}) & n \leq 0
 \end{array}
  \right.
  \end{equation}
  \end{widetext}
 for $\sigma<0$. In Eqs.(30) and (31), $\tau_n$ is defined by Eq.(21), whereas $H(\tau)$ is the step (Heaviside) function, i.e. $H(\tau)=0$ for $\tau<0$ and $H(\tau)=1$ for $\tau>0$.
 Like in the previous case, i.e. below the $\mathcal{PT}$ symmetry breaking, the wave function evolution is strongly asymmetric for reversal of the magnetic flux. As an example, in Fig.4 we show the numerically-computed evolution, both in real and momentum space, of the wave function corresponding to the intial particle at rest and fully delocalized on the ring, i.e. $c_n(0)=\delta_{n,0}$, for parameter values $V_0mR^2/ \hbar^2=0.02$, $\alpha=1$, and for  $\sigma=-0.003$ [Fig.4(a)] and $\sigma=0.003$ [Fig.4(b)]. The behavior of the (exact) numerically-computed wave function evolution in momentum space clearly reproduces the predictions based on the asymptotic (approximate) solutions as given by Eqs.(30) and (31). In particular, in the  $\sigma<0$ case [Fig.4(a)] the dynamics is frozen [$a_n(\tau) \simeq a_n(0)=\delta_{n,0}$], i.e. the potential $V(\varphi)$ appears to be invisible (like in Refs. \cite{P4,P5}) and the electromotive force does not increase anymore the angular momentum of the particle. Coversely, for $\sigma >0$ higher winding number states are generated owing to a sequence of LZ transitions, see Fig.4(b). Hence at the
  $\mathcal{PT}$-symmetry breaking point the LZ transitions are {\it unidirectional}.
  \par
 A striking effect, that we refer to as  {\it field-induced delayed transparency}, is the possibility for $\sigma<0$ to make the external potential $V(\varphi)$ "invisible" after some time delay $T$ from the initial time $\tau=0$ by application of a linearly growing magnetic flux. In other words, the particle motion is affected by the external potential $V(\varphi)$ up to the time $\tau=T$,  whereas for times $\tau>T$ the particle motion occurs as if the external potential $V$ were switched off (in spite it is still there). Such a counterintuitive effect 
 can be explained on the basis of {\it unidirectional} Zener tunneling between adjacent levels that enables to  freeze the particle motion in momentum space after some target time $T \geq 0$. In fact, let us assume that at time $\tau=0$ the particle is prepared in a rather arbitrary state with amplitude probabilities $c_n(0)$ in momentum space. Owing to the convergence of the series (16), one has $c_n(0) \rightarrow 0$ as $n \rightarrow \pm  \infty$. In practice, we may assume that $c_n(0) \simeq 0$ for $n \leq M$, where $M$ is some integer number (possibly negative and larger in absolute value). At time $\tau=0$, let us apply a linearly varying magnetic flux $f(\tau)=\sigma(\tau-\tau_0)$, where the parameter $\tau_0$ -to be determined- is the time at which the magnetic flux vanishes. Assuming $\sigma<0$, from Eqs.(19) -with $S_2=0$ and with $\tau$ replaced by  $\tau-\tau_0$ on the right hand side of the equations- it follows that, apart from the dynamical phase, the amplitude $c_n(\tau)$ is not affected by the external potential $V(\varphi)$ at times $\tau>\tau_0+\tau_{n-1}$ because Zener tunneling is prevented. Moreover, if $c_n(0)=0$ for $n \leq M$, for the unidirectionality of Zener tunneling it readily follows that $c_n(\tau) \simeq 0$ at any time $\tau>0$ for $n \leq M$. Hence the dynamics of the system is expected to freeze at times $\tau>\tau_{M}+\tau_0$. Physically, this means that the potential $V$ becomes "invisible" after a time delay 
  \begin{figure}[t]
\includegraphics[width=8.7cm]{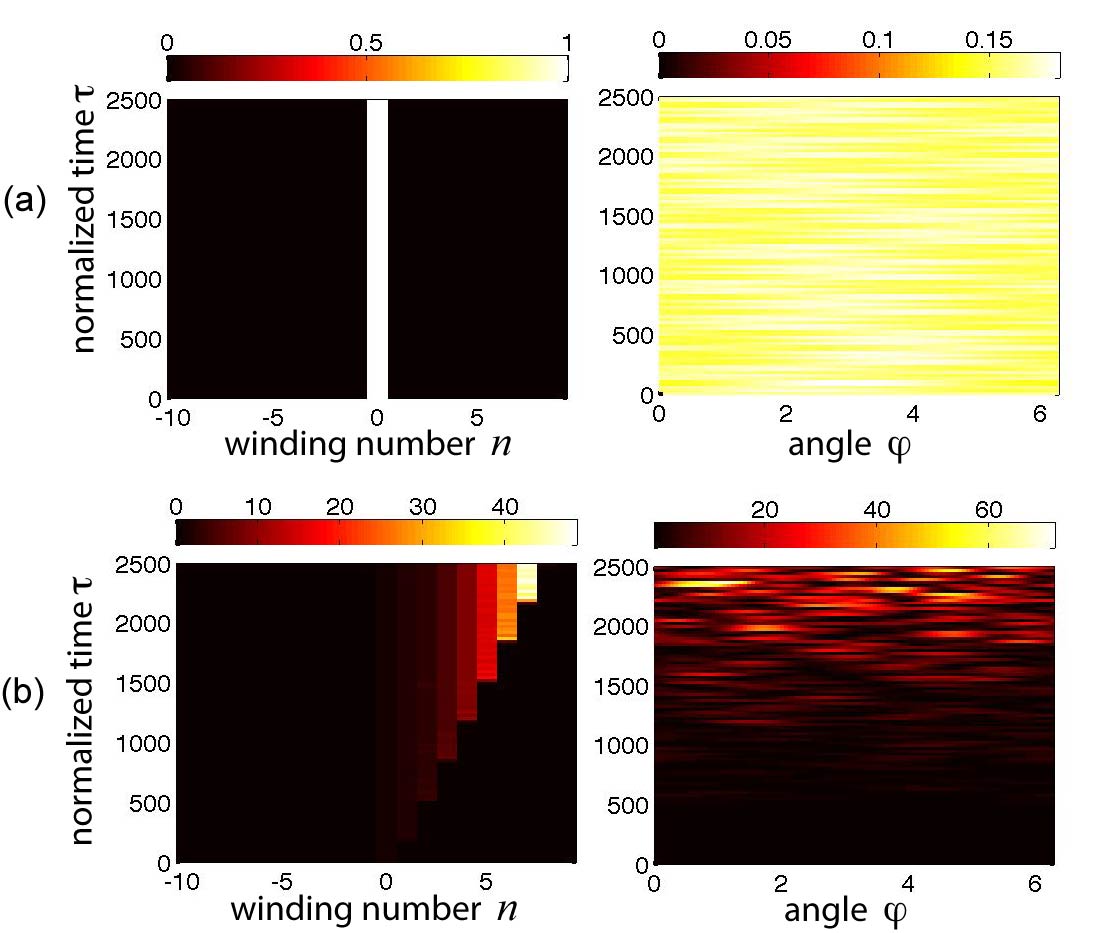}
\caption{(Color online). Same as Fig.3, but at the $\mathcal{PT}$-symmetry breaking point $(\alpha=1)$. In (a) $\sigma=-0.003$, in (b) $\sigma=0.003$. The amplitude of the external potential is $V_0mR^2/ \hbar^2=0.02$.}
\end{figure}
 \begin{equation}
 T=\tau_{M}+\tau_0=\frac{2M+1}{2 \sigma}+\tau_0.
 \end{equation}
 \begin{figure}[t]
\includegraphics[width=8.7cm]{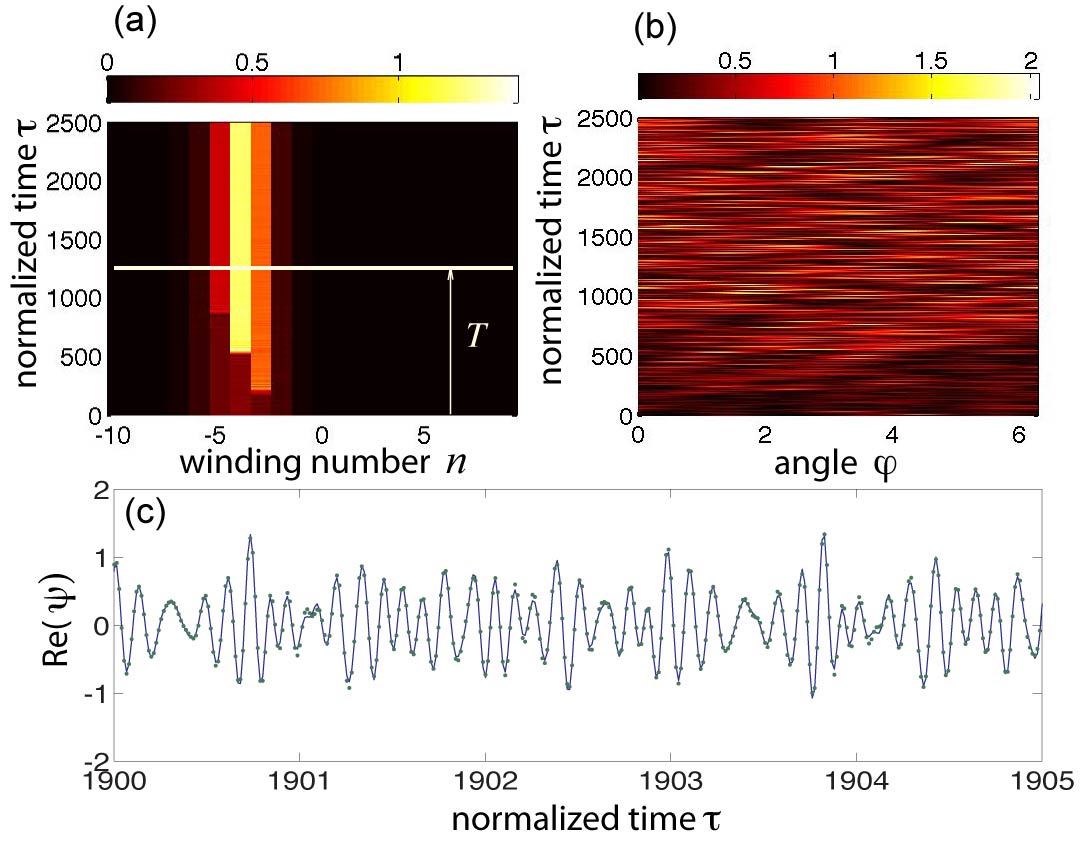}
\caption{(Color online). An example of delayed invisibility.  Wave function evolution, in momentum (a) and real space (b), for an initial Gaussian distribution in momentum space. Parameter values are given in the text. Note that at times $\tau > \simeq T$ the dynamics of occupation probabilities in momentum space is frozen. In (c) the evolution of the real part of the wave function $\psi(\varphi,t)$, at the angular position $\varphi=0$, is shown by the solid curve. The dotted curve, almost overlapped with the solid one, is the corresponding wave function evolution that one would observe by switching off the external potential $V$ at times $\tau>T$.}
\end{figure}
 i.e. at time $\tau>T$ the particle dynamics is not affected anymore by the external potential $V$. From Eq.(32) it follows that the delay $T$ can be made arbitrary by an appropriate choice of $\tau_0$. 
The occurrence of such a  delayed invisibility, predicted by the asymptotic analysis,  has been checked by direct numerical integration of Eqs.(19). An example of delayed invisibility is shown in Fig.5 for parameter values $\sigma=-0.003$, $\alpha=1$ and $mR^2 V_0 / \hbar^2=0.02$. As an initial condition, we choose a Gaussian distribution in momentum space, namely $c_n(0)= \mathcal{N} \exp[-(n+4)^2/9]$, where $\mathcal{N}$ is a normalization constant. Such a distribution has a negligible occupation amplitudes for $n < M \simeq -7$. Assuming $\sigma=-0.003$, to obtain a target delay of e.g. $T=1200$ according to Eq.(32) we take $\tau_0 \simeq -967 $. The numerically-computed evolution of the  wave function, both in momentum and real space, is shown in Figs.5(a) and (b), respectively. From Fig.5(a) it can be seen that the winding number occupation probabilities are frozen after a time $\tau \sim T$, where LZ transitions are forbidden. To check that the evolution of the wave function at times $\tau>T$ is not influenced anymore by the external potential $V$, i.e. that the external potential is effectively invisible at times $\tau>T$, we compared the wave function evolution in real space at times $\tau>T$ with that obtained by switching off the external potential at $\tau>T$, i.e. by letting $V(\varphi)=V_0 \exp(i \varphi)$ for $\tau<T$ and $V(\varphi)=0$ for $\tau>T$. As an example, in Fig.5(c) we show the behavior of the real part of the wave function at the azimuthal angle $\varphi=0$ over a time interval after $T$, as obtained in the two cases. Note the good overlap of the two curves, which indicates that the particle motion is effectively insensitive to the external potential.

\section{Conclusions}
Quantum mechanics in doubly-connected (ring) topologies in presence of a magnetic field, i.e. in the so-called quantum ring systems,  has long fascinated physicists, mainly because of  the manifestation of important physical effects such as the Aharonov-Bohm effect and persistent currents. In this work we extended the  theory of quantum rings by allowing for an external {\it non-Hermitian} potential. For a static magnetic flux, the quantum states of
the particle on the ring can be mapped onto the Bloch states of a complex crystal, and magnetic flux tuning enables to probe the spectral features of the 
complex crystal, including the appearance of exceptional points. For a time-varying (linearly-ramped) magnetic flux,  Zener tunneling among energy states is realized owing to the induced electromotive force on the ring. As compared to the Hermitian case, striking effects have been predicted to occur in the non-Hermitian case as a result of an asymmetric Zener tunneling. In particular, we discussed the possibility to observe delayed invisibility of an external potential at the $\mathcal{PT}$ symmetry breaking point. Hence  non-Hermitian quantum rings could provide a means to probe the spectral properties of complex crystals and to observe unusual phenomena, like delayed transparency.
The possibility to physically implement in a classical or quantum system the non-Hermitian quantum ring Hamiltonian (1)  remains an open question, which however goes beyond the scope of the present study. As briefly mentioned in Sec.II.A, possible physical systems where non-Hermitian quantum rings might be in principle realized include light propagation in twisted fibers \cite{fibra} or cold atoms in rotating annular traps \cite{BEC}. However,
for such systems the experimental and technological feasibility to implement a non-Hermitian potential, like the one considered in the present work [Eq.(10)], remains a rather challenging task. Different and experimentally more feasible physical implementations, for example based on propagation of optical pulses in recirculating fiber loops with gain and loss \cite{P6}, should be investigated. Finally, our analysis could be extended to study the dynamics of non-Hermitian quantum rings including a possible imaginary vector potential in the Schr\"{o}dinger equation \cite{Hatano}, in addition to the external potential.


\begin{thebibliography}{99}
\bibitem{note}
In the matter wave  realization of the Hamiltonian (1), discussed in Ref.\cite{BEC} and briefly mentioned in Sec.II.A, a linear increase in time of the magnetic flux can be realized by uniform acceleration of the condensate  around the annular trap. Similarly, in the optical realization of the Hamiltonian (1), discussed in Ref.\cite{fibra},  a  time-varying  magnetic flux can be realized by a non-uniform twist of the optical fiber along the propagation (axial) distance.
\end{thebibliography}
\end{document}